%% file: main.tex
\documentclass[article,nojss]{jss}
\usepackage{thumbpdf,lmodern}
\usepackage{framed} 
\usepackage{float}
\usepackage{enumitem}
\usepackage{comment}

\usepackage{xcolor}
\usepackage{amsfonts} 
\definecolor{orcidlogocol}{HTML}{A6CE39}
\usepackage{amsmath}
\newcommand{\indep}{\rotatebox[origin=c]{90}{$\models$}}

\newcommand{\fct}[1]{\code{#1()}}
\newcommand{\PM}{PM$_{2.5}$}

\newtheorem{definition}{Definition}
\newtheorem{assumption}{Assumption}

\author{Naeem Khoshnevis 
\\Harvard University
   \And Xiao Wu  
   \\Columbia University
   \And Danielle Braun  
   \\Harvard University}
\Plainauthor{Naeem Khoshnevis, Xiao Wu, Danielle Braun}

\title{CausalGPS: An R Package for Causal Inference With Continuous Exposures}
\Shorttitle{Weighting/Matching with Continuous Exposures}

\input{abstract}


\Address{
  Naeem Khoshnevis\\
  Research Computing and Data Services\\
  Harvard University\\
  1430 Massachusetts Ave, Cambridge, MA 02138\\
  E-mail: \email{nkhoshnevis@g.harvard.edu}\\
  \\
  Xiao Wu\\
  Department of Biostatistics\\
  Columbia University Mailman School of Public Health\\
  722 West 168th St., New York, NY 10032\\
  E-mail: \email{xw2892@cumc.columbia.edu}\\
  \\
  Danielle Braun\\
  Department of Biostatistics\\
  Harvard T.H. Chan School of Public Health \\
  Department of Data Science\\
  Dana-Farber Cancer Institute \\ 
  E-mail: \email{dbraun@mail.havard.edu}\\
}

\begin{document}

\input{content}


\newpage

\input{appendix}


\end{document}

%% file: abstract.tex
\Abstract{
Quantifying the causal effects of continuous exposures on outcomes of interest is critical for social, economic, health, and medical research. However, most existing software packages focus on binary exposures. 
We develop the \pkg{CausalGPS} \proglang{R} package that implements a collection of algorithms to provide algorithmic solutions for causal inference with continuous exposures. \pkg{CausalGPS} implements a causal inference workflow, with algorithms based on generalized propensity scores (GPS) as the core, extending propensity scores (the probability of a unit being exposed given pre-exposure covariates) from binary to continuous exposures. 
As the first step, the package implements efficient and flexible estimations of the GPS, allowing multiple user-specified modeling options. As the second step, the package provides two ways to adjust for confounding: weighting and matching, generating weighted and matched data sets, respectively. Lastly, the package provides built-in functions to fit flexible parametric, semi-parametric, or non-parametric regression models on the weighted or matched data to estimate the exposure-response function relating the outcome with the exposures. The computationally intensive tasks are implemented in \proglang{C++}, and efficient shared-memory parallelization is achieved by OpenMP API. This paper outlines the main components of the \pkg{CausalGPS} \proglang{R} package and demonstrates its application to assess the effect of long-term exposure to \PM\ on educational attainment using zip code-level data from the contiguous United States from 2000-2016.}

%% file: content.tex
\input{introduction}

\input{background_and_methods}

\input{package_overview}

\input{application_and_illustrations}

\input{conclusion}

\input{computational_details}

\input{acknowledgments}

\bibliography{refs}

%% file: introduction.tex
\section[Introduction]{Introduction} \label{sec:intro}

Studying causal relationships between a continuous exposure and an outcome of interest is a central consideration for scientists and policymakers in many disciplines.
\citet{robins2000marginal} and \citet{wu2018matching} have proposed new weighting and matching approaches, respectively, to handle continuous exposures in causal inference applications with observational data. While both authors provide tailored code for implementing the proposed approaches to their specific data applications, the barrier still exists for researchers to implement these generic statistical approaches for causal inference to broader data applications. 
Several other \proglang{R} packages currently exist, including \pkg{ctseff} \citep{kennedy2017non}, \pkg{independenceWeights} \citep{independenceWeights_2022}, \pkg{CBPS} \citep{CBPS_2022}, \pkg{twangContinous} \citep{twangContinuous_2021}. However, they tend to focus on individual weighting algorithms, offering limited options for user customization. In addition, none of these packages provide a unified workflow for implementing both weighting and matching approaches.

In this paper, we present generic and scalable statistical software for weighting and matching based on \cite{robins2000marginal} and \cite{wu2018matching}. Operationally, we develop the \pkg{CausalGPS} \proglang{R} package, implementing two sophisticated causal inference approaches, providing a handful of functionality that allows user specifications. When developing the package, we adhered to software engineering best practices to ensure efficiency, reliability, and modularity. This package's repository is hosted on GitHub, encouraging community contributions and transparent development.

\textbf{Software requirements:} \pkg{CausalGPS} works in conjunction with the \proglang{R} programming language and statistical software \citep{R_2023} and will run on any platform where \proglang{R} is installed (Windows, Unix, or Mac OS X). \pkg{CausalGPS} is available from the Comprehensive R Archive Network (CRAN) at \url{https://CRAN.R-project.org/package=CausalGPS}.  \pkg{CausalGPS} has been tested on the most recent version of \proglang{R}. 

%% file: background_and_methods.tex
\section{Background and methods} \label{methods}

\subsection{Basic concepts and notations}
\label{assumption}

Randomized controlled experiments are frequently regarded as the gold standard for evaluating causality \citep{harder2010propensity,imbens2015causal}. Nevertheless, there are numerous situations where it is impractical, unfeasible, or unethical to facilitate such randomized experiments. Consequently, scientists regularly rely on data from non-randomized, observational studies to address scientific questions. However, these studies are prone to confounding bias, a challenge not present in randomized experiments where the exposure assignment is randomized. In observational studies, the exposure assignment is not only unknown but often influenced by extraneous factors (i.e., potential confounders). \cite{rubin_1974} introduced confounding controls in observational studies within a potential outcomes framework.

We describe the potential outcomes framework in the context of continuous exposures. Formally, we use the following mathematical notation: Let $N$ denote the study sample size. For each unit, let $X$ denote the pre-exposure covariates; $E$ denote the observed continuous exposure; $Y$ denote the observed outcome.

\begin{definition} The potential outcomes are defined as a set of random variables, 
$Y(e) , \forall\ e \in [e^0,e^1]$, in which
$
Y = Y (E), \forall\ E \in [e^0,e^1].
$
\end{definition}

\begin{definition} The generalized propensity score (GPS) is the conditional density function of the exposure given  potential confounders
:
$
\mathbf{q}({x}) = \{f_{E\mid{X}} (e\mid{x}), \forall e \in [e^0,e^1]\}.
$
The single scores $q(e,{x})=f_{E\mid{X}} (e\mid{x})$ are  realizations of $\mathbf{q}({x})$ at exposure level $e$. 
\end{definition}
A key property of GPS is the covariate balancing property, that is, conditional on the GPS, the probability of receiving any level of exposure is independent of $X$ \citep{rosenbaum1983central}. By ensuring covariate balance, a pseudo-population can be created that mimics a randomized experiment \citep{desai2019alternative}.
Note that covariate balance is achieved by design for both measured and unmeasured covariates in randomized experiments, whereas the pseudo-population created by using GPS approaches in observational studies can achieve covariate balance only for measured covariates.

The potential outcomes framework allows us to 1) describe causal estimands of interests and their corresponding target population and 2) state causal identification assumptions transparently and explicitly. We follow the large body of literature on causal inference to state the following assumptions for causal identification.
\begin{assumption}[Consistency] $E =  e$ implies $Y=Y(e)$ for all units. 
\end{assumption}
\begin{assumption}[Overlap] For all values of potential confounders ${x}$, the density function of receiving any possible exposure level $e \in [e^0,e^1]$ is positive:
$
f(e\mid{x}) > 0 \ \text{for all} \ e, \ {x}.
$
\end{assumption}


\begin{assumption}[Weak Unconfoundedness] 
For any possible exposure level $e$, in which $e$ is continuous in the range $[e^0,e^1]$; 
$
E \ \indep \ Y(e) \ \mid \ {X}.
$
\end{assumption}

The three causal assumptions stated above, along with the mild smoothness conditions, allow us to identify and estimate the following causal estimand: the average causal exposure-response function (ERF) or also called the exposure-response curve (ERC) \citep{hirano2004propensity,kennedy2017non,wu2018matching}. 
\begin{definition}
The \textit{average causal ERF} is
$\mu(e) = \mathbb{E}[Y(e)]$, for all $e\in [e^0,e^1]$.
\end{definition}


Compared to other regression approaches, the main advantage of causal inference approaches is the separation of the design and analysis stages  \citep{rubin2001using,rubin2008objective,stuart2008best,imbens2015causal}. In the design stage, investigators design the study: 1) define the causal estimands and the target population, 2) implement a design-based method such as matching or weighting to construct a matched or weighted dataset, and 3) assess the quality of the design using metrics such as covariate balance. At the end of the design stage, investigators create a pseudo-population that mimics a randomized experiment without using the outcome information.  The analysis stage proceeds only after the completion of the design stage, during which outcome analysis is conducted on the pseudo-population.

\subsection{Causal inference workflow}
\label{workflow}
\begin{figure}[H]
    \centering
    \includegraphics[width=1.0\textwidth]{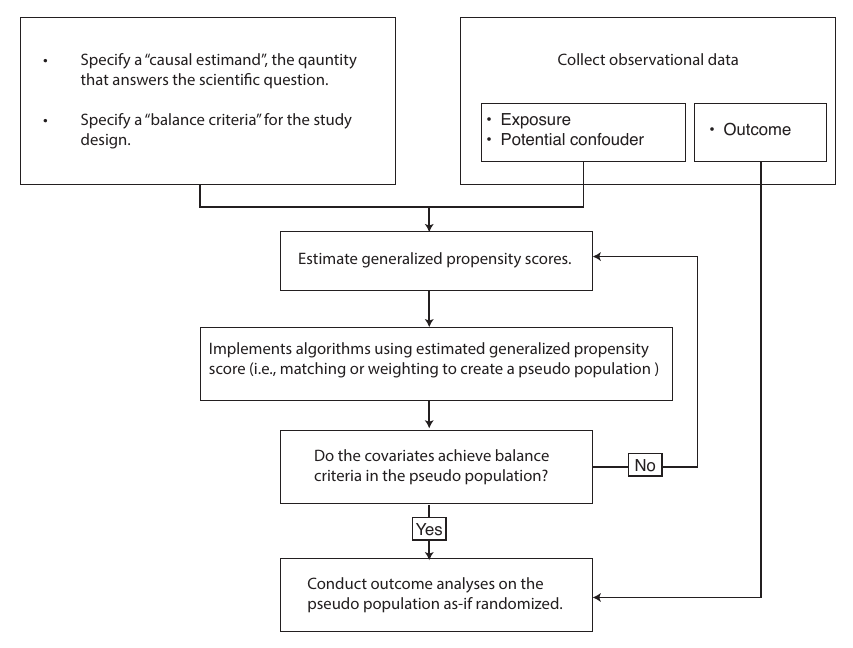}
    \caption{A Overview of the Causal Inference Workflow.}
    \label{fig:causal_workflow}
\end{figure}

We describe the causal inference workflow that follows the principle of separating the design and analysis stages. In summary, the workflow contains four steps: 1) estimation of the GPS, where exposure is regressed on the potential confounders; 2) implementation of the GPS weighting or matching approach to create a pseudo-population; 3) assessment of the quality of the pseudo-population in terms of covariate balance; 4) outcome analysis on the pseudo-population if covariate balance is achieved \citep{harder2010propensity}. Steps 1-3 belong to the design stage, and step 4 belongs to the analysis stage. An overview of the workflow is presented in Figure~\ref{fig:causal_workflow}. The primary focus of this paper is to introduce an \proglang{R} software package designed for implementing causal inference methods with continuous exposures. Therefore, we have chosen not to delve into the technical details of the different causal inference methods; however, we encourage readers seeking a deeper understanding of these methodologies to see \cite{robins2000marginal,zhu2015boosting,austin2018assessing,wu2018matching}.

%% file: package_overview.tex
\section{Package overview} \label{package_overview}

Following the workflow described in Section~\ref{methods}, the \pkg{CausalGPS} package's functionality can be divided into four major components: 1) estimating GPS values, 2) weighting/matching by GPS values to generate pseudo-populations, 3) assessing covariate balance on the pseudo-populations, and 4) outcome modeling on the pseudo-populations. To estimate GPS values, two methods are incorporated: a) a regression approach that assumes normally distributed residuals, hereafter referred to as \textit{normal}, and b) a kernel-based density estimation approach, hereafter referred to as \textit{kernel}. The GPS estimation depends on the internal hyperparameters used in the models. After obtaining estimated GPS values, the GPS weighting or matching algorithms are implemented to generate pseudo-populations. A covariate balance assessment based on absolute correlations between the exposure and observed pre-exposure covariates is applied to the generated pseudo-populations. The objective of assessing covariate balance is to determine how closely the distribution of observed pre-exposure covariates aligns across all levels of exposure. When covariate balance is not achieved, a stochastic search approach is implemented to tune different hyperparameters for the GPS model in a data-driven procedure and to operate covariate transforms to ensure covariate balance. For outcome modeling, \textit{parametric}, \textit{semi-parametric}, and \textit{non-parametric} approaches can be used to estimate the ERF on the generated pseudo-populations. 
The package is designed for optimal performance on shared memory systems. Parallelism at each stage is facilitated either by \proglang{R}'s \pkg{parallel} package \citep{R_2023} or by \pkg{Rcpp}'s OpenMP integration \citep{Rcpp_2011, Rcpp_2013, Rcpp_2018}. In the following, we discuss each step in more detail. 

\subsection{Estimating GPS values} \label{estimate_gps_values}

\begin{figure}[h]
    \centering
    \includegraphics[width=1.0\textwidth]{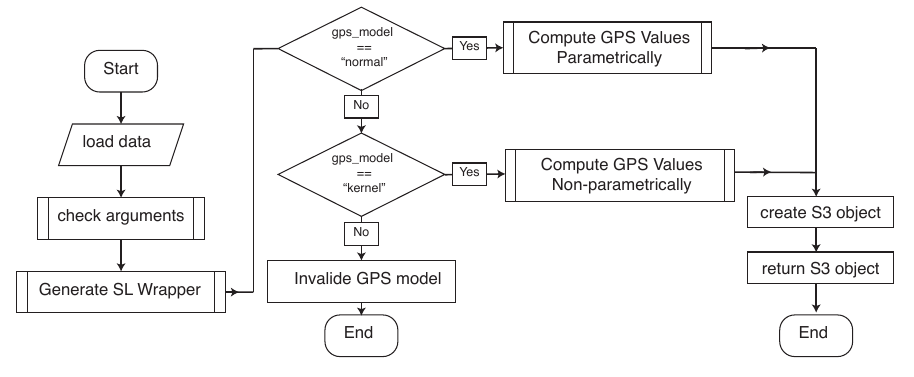}
    \caption{Workflow to estimate GPS values. Refer to the text for the input data and parameters. We generate a wrapper function, based on input hyperparameters, to be used by the \pkg{SuperLearner} package. The function returns an S3 \proglang{R} object.}
    \label{fig:estimate_gps}
\end{figure}

Both \textit{normal} and \textit{kernel} approaches for conditional density estimation of $f_{E|\mathbf{X}} (e | \mathbf{x})$ are implemented through the \fct{estimate\textunderscore gps} function to estimate GPS values (\code{gps\textunderscore density = normal} or \code{kernel}). The modeling assumptions for the \textit{normal} and \textit{kernel} approaches are different. Specifically, the \textit{normal} approach assumes that the conditional density of $E|\mathbf{X} = \mathbf{x}$ follows a normal distribution, whereas the kernel approach does not make parametric distributional assumptions on the conditional density and uses a non-parametric kernel smoother to estimate the conditional density of $E|\mathbf{X} = \mathbf{x}$. In either \textit{normal} or \textit{kernel} approaches, the \pkg{SuperLearner} \citep{SuperLearner_2021} package is used for generating an ensemble of machine learning models to estimate both the mean function $\mathbb{E}[E | \mathbf{C}]$ and the variance function $\mathbb{V}[E | \mathbf{C}]$. Figure~\ref{fig:estimate_gps} represents the workflow for estimating GPS values. A general syntax to run the function is: 
\begin{CodeChunk}
    \begin{CodeInput}
R> library("CausalGPS")
R> gps_obj <- estimate_gps(w,
                           c,
                           gps_density = "normal",
                           params = list(),
                           sl_lib = c("m_xgboost"),
                           nthread = 1,
                           ...) {
    \end{CodeInput}
    \end{CodeChunk}

where \code{w} requires inputs of a data frame containing observed continuous exposures, \code{c} requires inputs of a data frame containing observed (pre-exposure) covariates, and \code{gps_density} determines the model specifications and takes one of two options: \code{normal} or \code{kernel}. Hyperparameters to control internal libraries are passed through \code{params} input parameter. The \code{sl_lib} input parameters include the list of internal libraries that \pkg{SuperLearner} package uses in ensemble models. A wrapper function is generated to modify default parameters for these libraries.  The supported modified libraries are prefixed with \code{m\textunderscore}, and their supported hyperparameters are prefixed with predefined names. Table~\ref{tab:wrapper} shows supported modified wrappers. The wrappers can select hyperparameters randomly if a range of parameters is provided. The \code{nthread} parameter designates the number of cores the function can use in parallel processing. According to \proglang{R} \citep{R_2023} conventions, the ellipsis \code{"..."} is used to pass an arbitrary number of additional arguments.

\begin{table}
    \centering
    \caption{Available modified libraries for the \pkg{SuperLearner} package.}
    \begin{tabular}{cccp{6cm}}
        \hline
        Package name           & sl\textunderscore lib name                 & Parameter Prefix               & Supported Hyperparameters \\
        \hline
        \pkg{XGBoost}          & m\textunderscore xgboost                   & xgb\textunderscore             & nrounds, eta, max\textunderscore depth, min\textunderscore child\textunderscore weight, verbose \\
        \pkg{ranger}           & m\textunderscore ranger                    & rgr\textunderscore             & num.trees, write.forest, replace, verbose, family \\
        \hline
    \end{tabular}
    \label{tab:wrapper}
\end{table}

\subsection{Generating pseudo-population}

The pseudo-population refers to a data set that was either weighted or matched to make the distribution of pre-exposure covariates align across different exposure levels, that is, achieving covariate balance. A more in-depth explanation of covariate balance can be found in \citet{wu2018matching}. It is noteworthy that ensuring covariate balance across all exposure levels in observational data can be intricate, and imbalances might still arise even after the data has been weighted or matched.

\begin{figure}[h]
    \centering
    \includegraphics[width=1.0\textwidth]{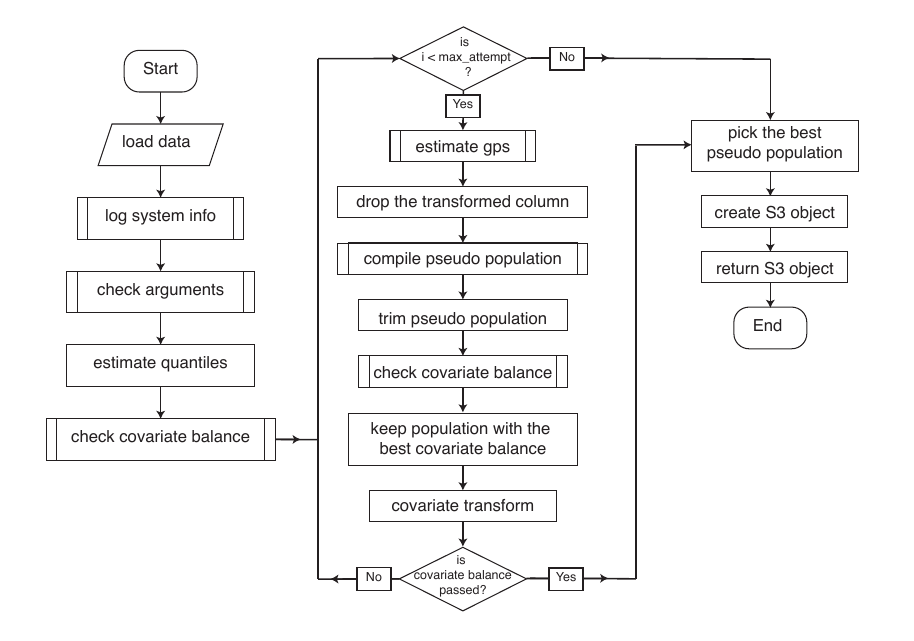}
    \caption{Workflow to generate pseudo-population. Refer to the text for the input data and parameters. The function attempts to achieve covariate balance by tuning hyperparameters, transforming covariates, and trimming data. The function returns an S3 \proglang{R} object. }
    \label{fig:generate_pseudo_pop}
\end{figure}

Firstly, the degree to which we can achieve covariate balance is heavily influenced by the estimated GPS values. Another strategy to improve covariate balance involves trimming data points with extreme exposure values, thereby focusing on data regions with substantial overlapping exposures. This process is visualized in Figure~\ref{fig:generate_pseudo_pop}, illustrating the iterative method to create a pseudo-population. Our \code{generate_pseudo_pop} function integrates hyperparameter tuning, covariate transformations, and exposure (and GPS) trimming to optimize the balance in the pseudo-populations.

A general syntax to run the function is:
\begin{CodeChunk}
\begin{CodeInput}
R> library("CausalGPS")
R> pspop_obj <- generate_pseudo_pop(w,
                                    c,
                                    ci_appr,
                                    gps_density = "normal",
                                    use_cov_transform = FALSE,
                                    transformers = list("pow2","pow3"),
                                    bin_seq = NULL,
                                    exposure_trim_qtls = c(0.01, 0.99),
                                    gps_trim_qtls = c(0.0, 1.0),
                                    params = list(),
                                    sl_lib = c("m_xgboost"),
                                    nthread = 1,
                                    include_original_data = FALSE,
                                    gps_obj = NULL,
                                    ...){

\end{CodeInput}
\end{CodeChunk}
where \code{w}, \code{c}, and \code{gps\textunderscore density} correspond to the parameters found in the function \fct{estimate\textunderscore gps} detailed in Section~\ref{estimate_gps_values}. \code{ci\textunderscore appr}, stands for the causal inference approach. The currently available options are \code{weighting} and \code{matching}. \code{bin\textunderscore seq} represents the sequence of exposure points at which the ERF will be evaluated. \code{exposure\textunderscore trim\textunderscore qtls} is used to trim out the data samples near the support boundaries of the exposure distribution, and \code{gps\textunderscore trim\textunderscore qtls} is used to trim out the data samples based on the GPS values. Since the input data might undergo various trimming procedures, each row is assigned a distinct ID column (\code{id}) to facilitate any necessary data merge.

Setting \code{use\textunderscore cov\textunderscore transform = TRUE} activates an iterative approach to obtain GPS values under different hyperparameters and covariates transformations, and choose the generated pseudo-population that achieves acceptable covariate balance. At each iteration, we apply a transformer to the covariate with the highest absolute correlation with the exposure (i.e., the most imbalanced covariate). Each transformer is a univariate function and can be defined by the user. We implement two transformers, \code{pow2} and \code{pow3}, which represent the power of two and three transformations (transforming $X$ into $X^2$ and $X^3$, respectively). 
Additional transformers can be given as a list of univariate functions to the input parameter \code{transformers}. In every subsequent iteration, the corresponding transformer from the \code{transformers} list will be applied to the most imbalanced covariate from the previous iteration. Within a single iteration, only one covariate will be subject to transformation. Note that the covariate transformation is only used to estimate GPS values, whereas the covariate balance, represented as absolute correlations, is still computed between the exposure and the untransformed scale of the covariates. This iterative approach for choosing the pseudo-population with minimized absolute correlations can be time-consuming. Users can limit the total number of iterations using the input parameter \code{max\textunderscore attempt}. Since hyperparameters of a GPS model (in estimating GPS values) are based on the list of values provided by users, the parameters used in each new iteration are different, and not necessarily better than, the prior iteration, resulting in different degrees of covariate balance. The resulting covariate balance from each iteration will be automatically compared to a user-specified threshold to determine whether the covariate balance is acceptable. The covariate balance threshold (\code{covar\textunderscore bl\textunderscore trs}) can be applied to the \code{"maximal"}, \code{"median"}, and \code{"mean"} values of the absolute correlations of each covariate with the exposure variable. The search process concludes either upon achieving acceptable covariate balance or upon reaching the maximum number of iterations (\code{max_attempt}), whichever comes first. The input parameter \code{include\textunderscore original\textunderscore data = TRUE} is used to include original data (without any trimming or transformation) in the output object. If the user provides pre-trained GPS models via the input parameter \code{gps\textunderscore object}, the derived GPS values will be used in weighting or matching to generate the pseudo-population, and the GPS estimation step will be skipped.

The following parameters are specific to the implemented causal inference approach. For the \code{matching} approach, the following parameters are required:

\begin{itemize}
    \item \code{dist\textunderscore measure:} Distance measure to identify the neighborhood of units. The available option is \code{l1} for Manhattan distance.
    \item \code{delta\textunderscore n:} Caliper size on the continuous exposure. 
    \item \code{scale:} A scale parameter to control the relative weight attributed to the distance measures of the exposure values versus GPS values. 
    \item \code{covar\textunderscore balance\textunderscore method}: The covariate balance method. The available option is \code{absolute}, which calculates the absolute correlations.
    \item \code{covar\textunderscore balance\textunderscore trs:} The covariate balance threshold value (suggested value is 0.1).
    \item \code{covar\textunderscore balance\textunderscore trs\textunderscore type:} The covariate balance type. Available options include \code{"mean"}, \code{"median"}, and \code{"maximal"}. 
\end{itemize}

For the \code{weighting} approach, only the last three parameters are required.

\subsection{Modeling outcomes for causal inference}

After generating the balanced pseudo-population, the next step is to fit outcome models and then estimate the ERF. In the \pkg{CausalGPS} package, three outcome models are provided to estimate the ERF, allowing parametric, semi-parametric, and non-parametric models.
\begin{enumerate}
    \item The function \fct{estimate\textunderscore pmetric\textunderscore erf} estimates the coefficient between the outcome and the exposure, using a parametric regression model. By default, the function calls the \textbf{gnm} library to implement generalized nonlinear models.
    \item The function \fct{estimate\textunderscore semipmetric\textunderscore erf} estimates the smoothed ERF using a generalized additive model with splines. By default, the function calls the \textbf{gam} library to implement generalized additive models.
    \item The function \fct{estimate\textunderscore npmetric\textunderscore erf} estimates the smoothed ERF using a non-parametric kernel smoothing approach. By default, the function calls the \textbf{locpol} library to implement local polynomial regression models. We use a data-driven bandwidth selection for kernels.
\end{enumerate}

\subsection{Setting up logger}

The package uses \pkg{logger} \proglang{R} package as a logging infrastructure \citep{logger}. Users can select different log levels, including \code{TRACE}, \code{DEBUG}, and \code{INFO} (default). The following command can be used to set up logging parameters.
\begin{CodeChunk}
\begin{CodeInput}
R> set_logger(logger_file_path = "CausalGPS.log", logger_level = "INFO")
\end{CodeInput}
\end{CodeChunk}
where \code{logger\textunderscore file\textunderscore path} is the file path of the log file and \code{logger\textunderscore level} changes the logging level. Use the \fct{get\textunderscore logger} function to see the current parameters. 

\subsection{Efficiency, reliability, and modularity}

We emphasized the orthogonality of each component, allowing for modular design and ensuring each component operates independently. To ensure robustness and reliability, comprehensive unit testing was performed on each function. We supplemented unit testing with functional testing to evaluate the package's end-to-end functionality in real-world scenarios. For the user's benefit, we have incorporated detailed documentation for each function, and for developers, a comprehensive guide is available, facilitating seamless collaboration within the open-source community. Continuous integration is crucial in our development workflow, promoting regular code integration and ensuring consistent quality. Performance optimization was a key focus; computationally intensive modules were implemented in \proglang{C++} to boost execution speed. To ensure efficient scalability, the package can leverage multiple cores when available. The package is available on CRAN for user access and on GitHub for developers seeking to collaborate or extend its functionalities. Proactive maintenance ensures the package remains updated and relevant. To enhance user experience, all input parameters undergo checks, and informative messages guide the user throughout the processes. Additionally, we have implemented a logging mechanism to record essential internal operations, parameter values, and any randomized values. To ensure reproducibility across runs, users can set a seed value, ensuring consistent results. Furthermore, we have adhered to a predefined styling guide for consistent code presentation, and we have prioritized efficient implementation throughout the package's development.

%% file: application_and_illustrations.tex
\section{Application and illustrations}

The implemented weighting and matching approaches in the \pkg{CausalGPS} package can be applied to a wide range of observational studies with continuous exposures. As highlighted in Section~\ref{package_overview}, the \pkg{CausalGPS} package distinguishes between the design and analysis stages. In the design stage, where we initially identify causal estimands and the target population, followed by applying a design-based technique like weighting or matching, only exposures and pre-exposure covariates are needed for the input. Subsequently, we assess the design quality by checking the covariate balance. 
In the analysis stage, the outcome data is required.

We illustrate and discuss the main functionality and performance of the \pkg{CausalGPS} package using an air pollution data application. Before digging into the data set:

The following command installs the \pkg{CausalGPS} package. 
\begin{CodeChunk}
\begin{CodeInput}
R> install.packages("CausalGPS")
\end{CodeInput}
\end{CodeChunk}
The following command loads the \pkg{CausalGPS} package in the working space. 
\begin{CodeChunk}
\begin{CodeInput}
R> library("CausalGPS")
\end{CodeInput}
\end{CodeChunk}
\subsection{Data}

For illustrative purposes, we use publicly available air pollution data for each zip code across the contiguous United States from 2000 to 2016. The data set includes 580,244 zip code-years for 34,928  zip codes \citep{DVN_5XBJBM_2023}. The goal is to evaluate the causal relationship between long-term exposure to \PM\  and educational attainment. It is important to note that the entire example is intended solely to demonstrate the package's functionality; thus, the results should not be used for alternative objectives.  

Table~\ref{tab:datasummary} summarizes a list of variables collected for this study. The unit of this study is zip code-year. The \textit{Education rate} refers to the percentage of people over 65 years old who did not finish high school in each zip code-year. Socio-economic determinants, such as median household income, median house value, and poverty, offer insight into the economic conditions of the regions. Demographic information, such as the percentage of Hispanic and Black populations, provides a glimpse into the ethnic composition of each area. Meteorological variables, such as summer and winter min/max temperatures and relative humidity, are also included. Behavior risk factors include the smoke rate, corresponding to the fraction of the population who has ever smoked. 


\begin{table}[t]
    \centering
    \caption{Baseline characteristics of the data illustration by zip code across 2000-2016.}
    \begin{tabular}{lllllccc}
        \hline
        \multicolumn{4}{l}{Variables}                    & min     & median   & max     \\\hline
        \multicolumn{3}{l}{\citet{NASA_pm25}}    &       &         &          &         \\
                     & \multicolumn{3}{l}{\PM\ ($\mu g/m^3$)}    &  0.0078  &   9.7364  & 30.9249  \\
        \multicolumn{3}{l}{\citet{Walker_2021}}    &        &                  &         &  \\
                     & \multicolumn{3}{l}{Poverty}    &  0  &   0.0810  & 1.0  \\  
                     & \multicolumn{3}{l}{Race/ethnicity (\%)}        \\  
                     &        & \multicolumn{2}{l}{Hispanic}        & 0     & 0.2688 & 1.0         \\
                     &        & \multicolumn{2}{l}{Black}           & 0     & 0.0141 & 1.0         \\
                     & \multicolumn{3}{l}{Education rate}    &  0  &   0.2552 & 1.0  \\ 
                     & \multicolumn{3}{l}{Household Income (\$)}    &  0  &   44380 & 250001  \\ 
                     & \multicolumn{3}{l}{Median House Value (\$)}    &  0  &   118300.00 & 2000001.00   \\ 
                     & \multicolumn{3}{l}{Population Density}    &  0  &   139.7 & 153867.7  \\ 
        \multicolumn{3}{l}{\citet{CDC_BRFSS}}   &         &         &          &         \\
                     & \multicolumn{3}{l}{Mean Body Mass Index (kg/$m^2$)}    &  20.70  &   27.51  & 43.07  \\
                     & \multicolumn{3}{l}{Smoke rate (\%)}        &  0  &   0.47  & 1  \\ 
        \multicolumn{3}{l}{\citet{Abatzoglou_2013_IJC}}   &         &         &          &         \\
                     & \multicolumn{3}{l}{Mean Temperature ($K$)}        \\  
                     &        & \multicolumn{2}{l}{Summer\ time}    & 287.96     & 302.43 & 317.07         \\  
                     &        & \multicolumn{2}{l}{Winter\ time}    & 260.12     & 279.70 & 300.64         \\ 
                     & \multicolumn{3}{l}{Relative Humidity (\%)}        \\
                     &        & \multicolumn{2}{l}{Summer\ time}    &  22.35     & 91.37  &  100        \\  
                     &        & \multicolumn{2}{l}{Winter\ time}    &  35.79     & 87.23  &  100       \\
    \hline
    \end{tabular}
    \label{tab:datasummary}
\end{table}



\subsection{Illustrative Example}

Examples for running each function individually are provided in the \pkg{CausalGPS} package (see the vignettes, functions' documentation, and unit tests for more details). In this section, we choose one example that covers frequently used functionality throughout three different parameter specifications. We also discuss the parameter selection process and its associated results. 

In this example, we estimate the ERF to quantify the effect of exposure to annual \PM\ on educational attainment by implementing the GPS matching approach. In estimating the GPS, we apply the \textit{normal} approach for conditional density estimation and search available parameters to attain acceptable covariate balance. 
The primary parameters for this search are the \code{scale} and \code{delta_n}. The \code{scale} indicates the emphasis placed on the GPS when selecting the nearest neighbor, while \code{delta_n} represents the caliper size. In this demonstration, we set \code{scale = 1} (allocating the full emphasis to the GPS) and execute the iterative approach for  searching optimal \code{delta_n}. Based on \citet{wu2018matching}, we choose a search range of $[0.1, 2.4]$ with grids of 0.1. 
To help achieve improved covariate balance, the following four additional strategies were built into the package and can be deployed individually or jointly.

\begin{enumerate}
    \item Limit the analyses to the common support of exposure range using \code{exposure\textunderscore trim\textunderscore qtls}.
    \item Limit the analyses to the common support range of GPS values, as specified by \code{gps\textunderscore trim\textunderscore qtls}.
    \item Adjust the hyperparameters for ensemble machine learning models, using \code{params}.
    \item Opt for covariate transformations, using \code{transformers}.
\end{enumerate}

\begin{figure}[h]
    \centering
    \includegraphics[width=1\textwidth]{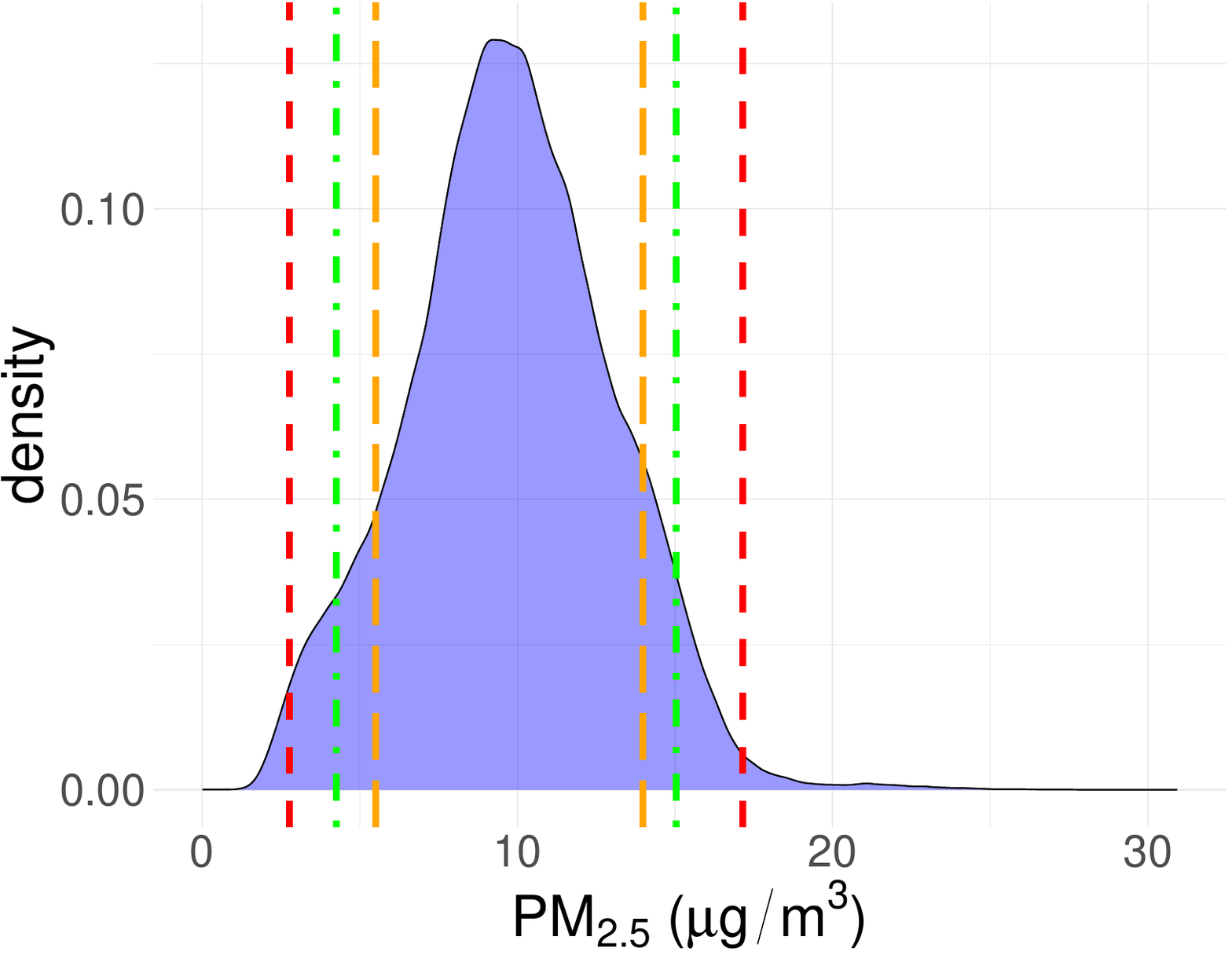}
    \caption{Density plot of \PM\ shows quantiles at (1st, 99th), (5th, 95th), and (10th, 90th). These quantiles are depicted using red dashed, green dot-dashed, and orange long-dashed lines, respectively.}
    \label{fig:exp_density}
\end{figure}

We showcase three distinct specifications of the parameter \code{exposure\textunderscore trim\textunderscore qtls}, varying the exposure trimming to the 1st and 99th percentiles, 5th and 95th percentiles, and 10th and 90th percentiles, respectively, out of concern for the positivity assumption. Figure~\ref{fig:exp_density} represents the density of \PM\ and trim quantiles.

The following code shows the function call and parameters for generating the pseudo-population.
\begin{CodeChunk}
\begin{CodeInput}
R> set.seed(249)
R> ps_pop_obj <- generate_pseudo_pop(data[, c("id", "pm25")],
                                     data[, c("id", confounders)],
                                     ci_appr = "matching",
                                     gps_density = "normal",
                                     bin_seq = NULL,
                                     exposure_trim_qtls = c(p_q1,
                                                            p_q2),
                                     use_cov_transform = TRUE,
                                     params = list(xgb_max_depth = c(3,4,5),
                                                   xgb_nrounds = seq(10, 40, 1)),
                                     sl_lib = c("m_xgboost"),
                                     nthread = 10,
                                     covar_bl_method = "absolute",
                                     covar_bl_trs = 0.1,
                                     covar_bl_trs_type= "maximal",
                                     max_attempt = 10,
                                     dist_measure = "l1",
                                     delta_n = delta_n,
                                     scale = 1)
\end{CodeInput}
\end{CodeChunk}
When \code{bin_seq = NULL}, the default values for \code{bin_seq} will be used, which is
\begin{CodeChunk}
\begin{CodeInput}
R> seq(min(w)+delta_n/2,max(w), by=delta_n)
\end{CodeInput}
\end{CodeChunk}
We limit the maximum number of iterations to 10 for each execution (see the running time of each execution in Table~\ref{tab:run_results}). Increasing the maximum number of iterations would allow a search for a larger parametric space of GPS models yet could potentially increase running time substantially.

Table~\ref{tab:run_results} presents the results under three specifications of exposure trimming, highlighting the resulting optimal caliper size, the maximal absolute covariate balance, the number of iterations, and the time taken for each execution. The results indicate that achieving covariate balance becomes more feasible upon excluding data points near the boundary of the exposure range. This can be attributed to the propensity of data points with extreme exposure values to be non-overlapped with other data points, thus complicating the process of identifying the nearest neighbor match. An alternative strategy to resolve the data non-overlapping issue is to exclude data points with extreme GPS values, which was built into the function \fct{generate\textunderscore pseudo\textunderscore pop} by the input parameter \code{gps\textunderscore trim\textunderscore qtls}. However, excluding data points from the analyses deviates the data distribution of the pseudo-population from the original population, thereby altering the definition of the causal estimand. It is imperative for users to interpret their results in the context of the target population, bearing in mind the specific causal estimand of interest.


\begin{table}[h]
\centering
\caption{Computational Results}
\begin{tabular}{cccccc}
\hline

Example & Trimming & Optimal & Covariate & Number of & Wall Clock\\
Name &  \code{c(p\_q1,p\_q2)} & \code{delta_n} & Balance &  Iterations &  Time (s) \\

\hline
\code{matching_1} & (1th, 99th) & 1.7 & 0.198 & 10 & 1219.37 \\
\code{matching_2} & (5th, 95th) & 0.9 & 0.123 & 10 & 1264.66 \\
\code{matching_3} & (10th, 90th) & 1.9 & 0.094 & 3 & 287.32 \\

\hline
\end{tabular}
\label{tab:run_results}
\end{table}

The output of \fct{generate\textunderscore pseudo\textunderscore pop}, \code{ps_pop_obj}, is a \code{S3} object that includes parameters and data to understand the process. Some of the main fields are:

\begin{itemize}[noitemsep, topsep=0pt]
    \item \code{params}: A list of input parameters.
    \item \code{pseudo_pop}: The matched pseudo-population.
    \item \code{adjusted_corr_results}: Covariate balance scores for the matched data.
    \item \code{original_corr_results}: Covariate balance scores for the original data. 
    \item \code{passed_covar_test}: A logical variable that represents if the covariate balance requirements are satisfied. 
    \item \code{best_gps_used_params}: Hyperparameters that are used for generating the selected pseudo-population.  
\end{itemize}

\begin{figure}[H]
    \centering
    \includegraphics[width=1\textwidth]{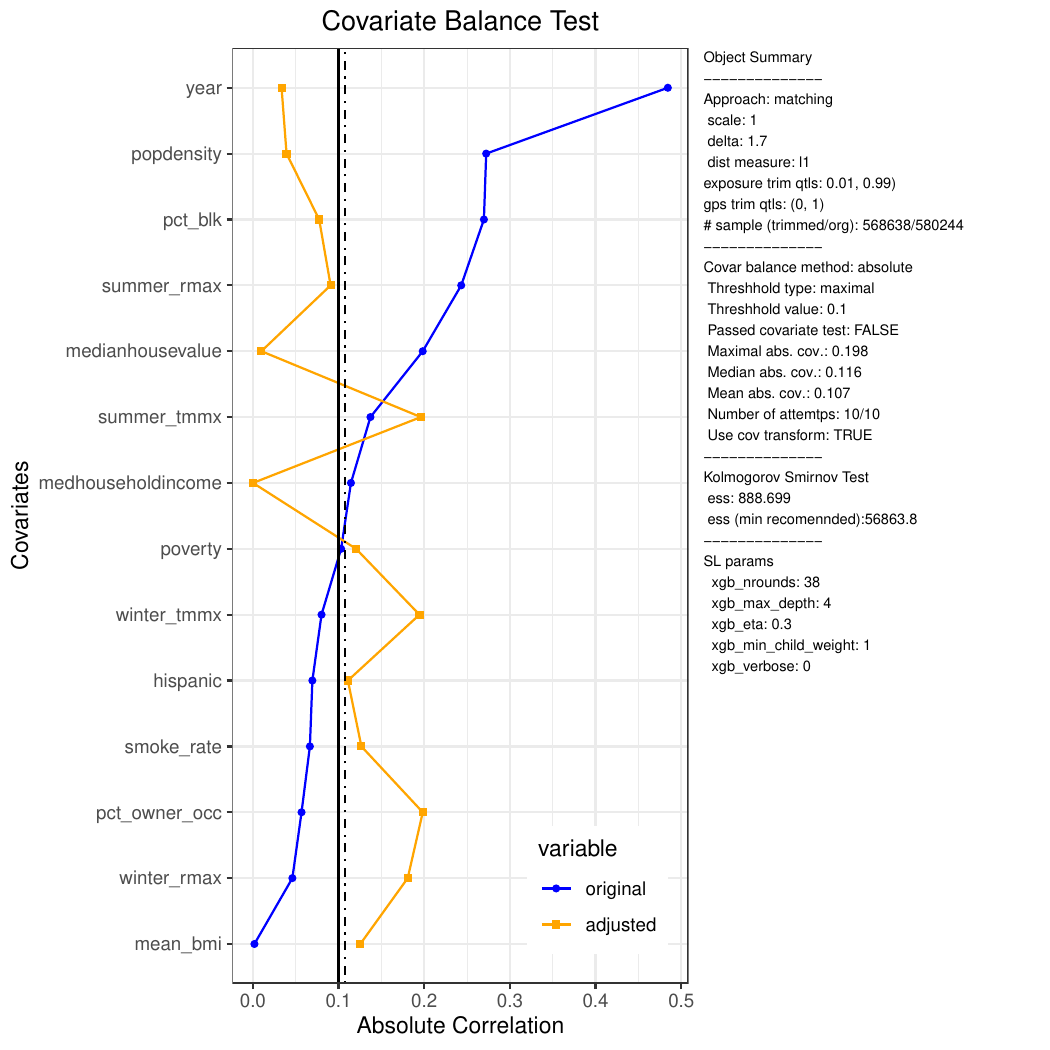}
    \caption{Covariate balance plot for \code{matching\textunderscore 1}. The covariates are sorted according to their covariate balance score in the original data. The solid vertical line represents the covariate balance threshold (\code{covar\textunderscore bl\textunderscore trs}), and the dashed vertical line is the mean absolute correlation value for the matched data. The object summary is included on the right side.}
    \label{fig:matching_1}
\end{figure}

\begin{figure}[H]
    \centering
    \includegraphics[width=1\textwidth]{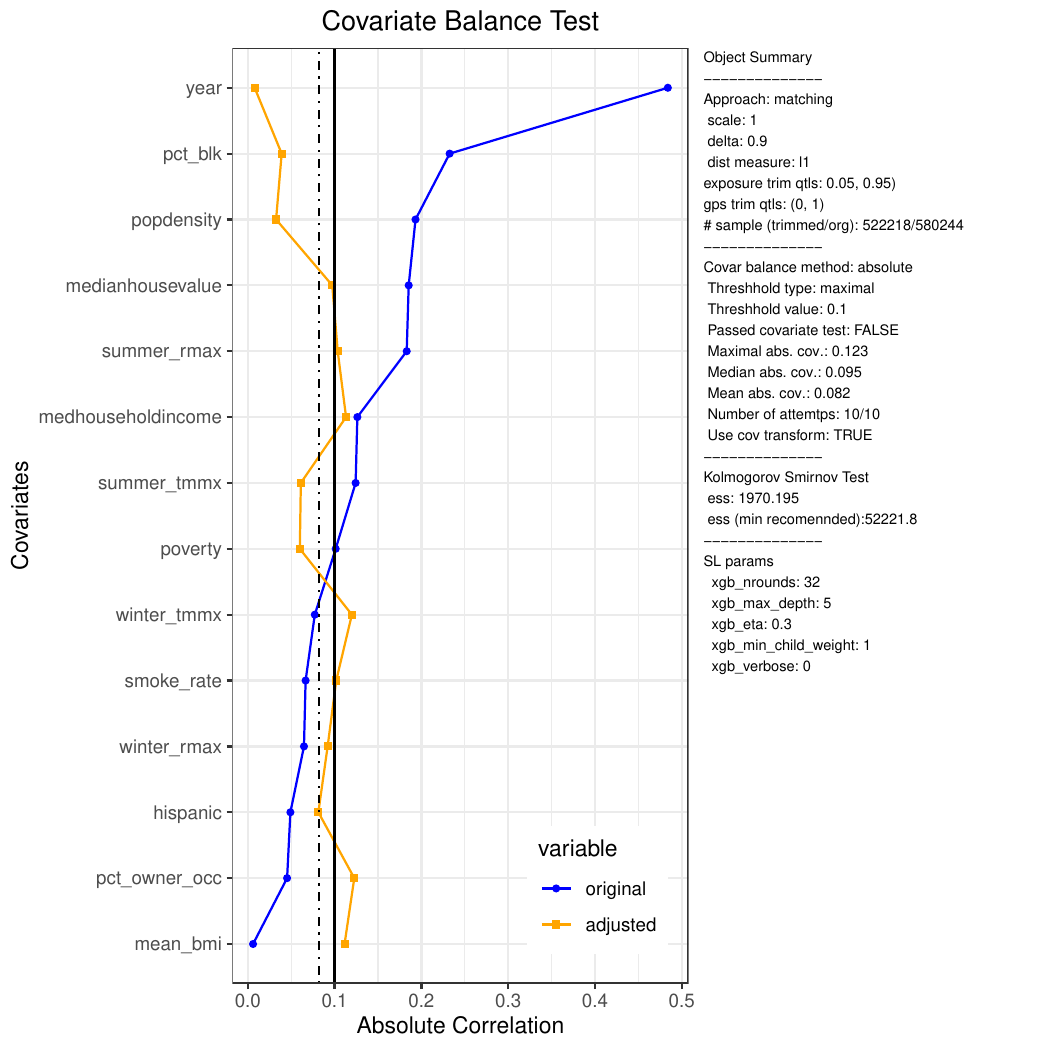}
    \caption{Covariate balance plot for \code{matching\textunderscore 2}. The covariates are sorted according to their covariate balance score in the original data. The solid vertical line represents the covariate balance threshold (\code{covar\textunderscore bl\textunderscore trs}), and the dashed vertical line is the mean absolute correlation value for the matched data. The object summary is included on the right side.}
    \label{fig:matching_2}
\end{figure}

\begin{figure}[H]
    \centering
    \includegraphics[width=1\textwidth]{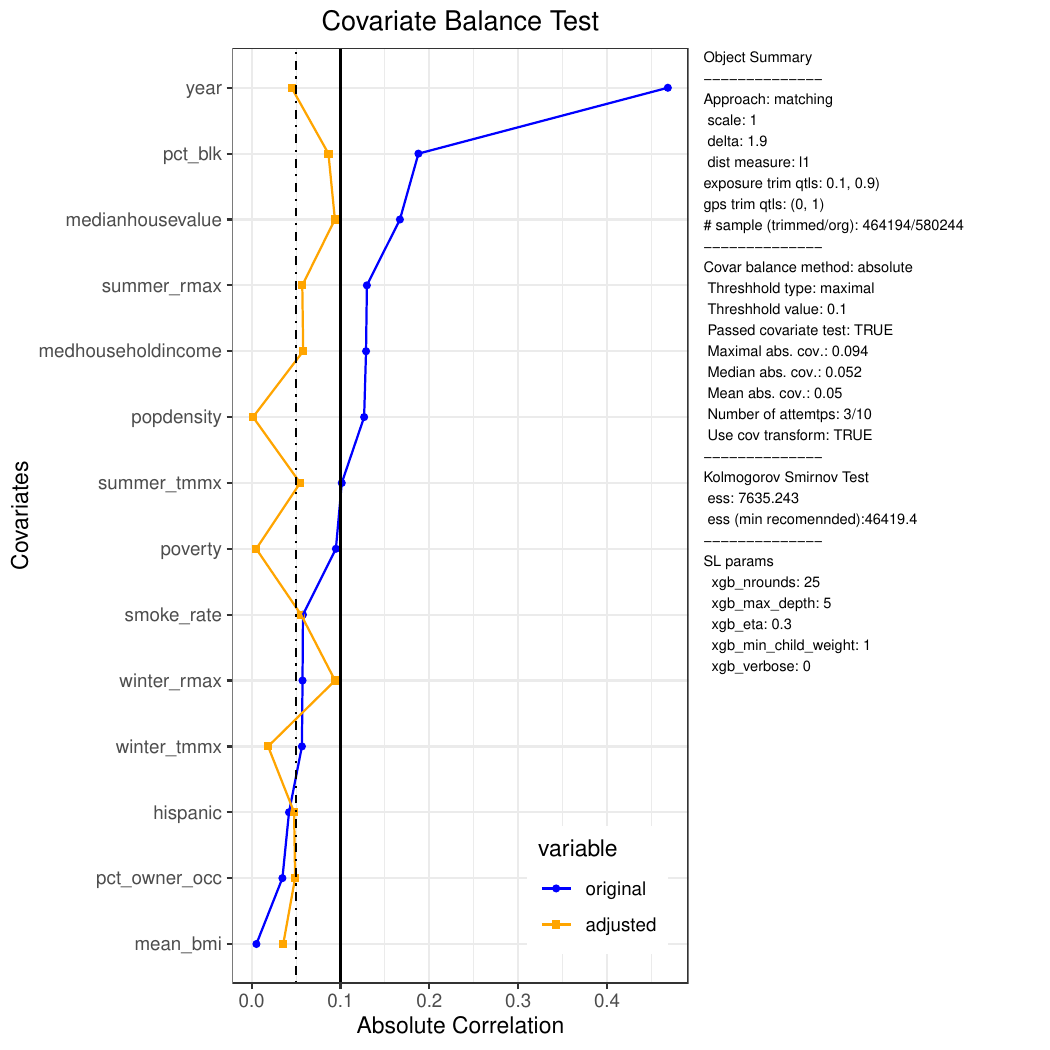}
    \caption{Covariate balance plot for for \code{matching\textunderscore 3}. The covariates are sorted according to their covariate balance score in the original data. The solid vertical line represents the covariate balance threshold (\code{covar\textunderscore bl\textunderscore trs}), and the dashed vertical line is the mean absolute correlation value for the matched data. The object summary is included on the right side.}
    \label{fig:matching_3}
\end{figure}

For a full list of items in the \code{S3} object \code{ps_pop_obj}, refer to the function help. 
\begin{CodeChunk}
\begin{CodeInput}
R> ?CausalGPS::generate_pseudo_pop
\end{CodeInput}
\end{CodeChunk}  
A summary of the covariate balance before and after matching can be obtained using the following function \fct{summary}. 
\begin{CodeChunk}
\begin{CodeInput}
R> summary(ps_pop_obj)
\end{CodeInput}
\end{CodeChunk}  
The package also provides a visualization tool for the \code{S3} object \code{ps_pop_obj}, allowing users to get a quick diagnosis of the matched data set. The plotting function offers an additional parameter, \code{include_details}. Setting this to \code{TRUE} will incorporate all essential details about the \code{S3} object \code{ps_pop_obj} on the resulting figure. 
\begin{CodeChunk}
\begin{CodeInput}
R> plot(ps_pop_obj, include_details = TRUE)
\end{CodeInput}
\end{CodeChunk}
Figures~\ref{fig:matching_1}-\ref{fig:matching_3} show outputs of the function \fct{plot} under three specifications of \code{exposure\textunderscore trim\textunderscore qtls} (i.e., \code{matching_1}, \code{matching_2}, and \code{matching_3}, respectively). Details of all parameters and diagnoses under each specification are presented on the right-hand side of the figure. By setting the covariate balance threshold \code{covar\textunderscore balance\textunderscore trs = 0.1}, and \code{covar\textunderscore balance\textunderscore trs\textunderscore type = "maximal"}, we require every single covariate maintain an absolute correlation $< 0.1$ with the exposure after matching. Only \code{matching_3} achieves the acceptable covariate balance. In general, setting \code{covar\textunderscore balance\textunderscore trs\textunderscore type = "maximal"} gives a more stringent constraint to covariate balance compared to \code{covar\textunderscore balance\textunderscore trs\textunderscore type = "mean"}, since the latter one only requires the mean value of absolute correlations for the matched data to be $< 0.1$, while certain covariates can be correlated with the exposure with an absolute correlation $\geq 0.1$, a condition that is satisfied by \code{matching_2}. Nevertheless, achieving more controls on covariate balance might require searching through a larger parameter space of GPS models, which could be time-consuming, and trimming more data points. Decisions regarding the covariate balance threshold, data trimming, and allowed running time will be left to users, who will make these determinations based on practical considerations specific to their data applications.

\begin{figure}[H]
    \centering
    \includegraphics[width=1\textwidth]{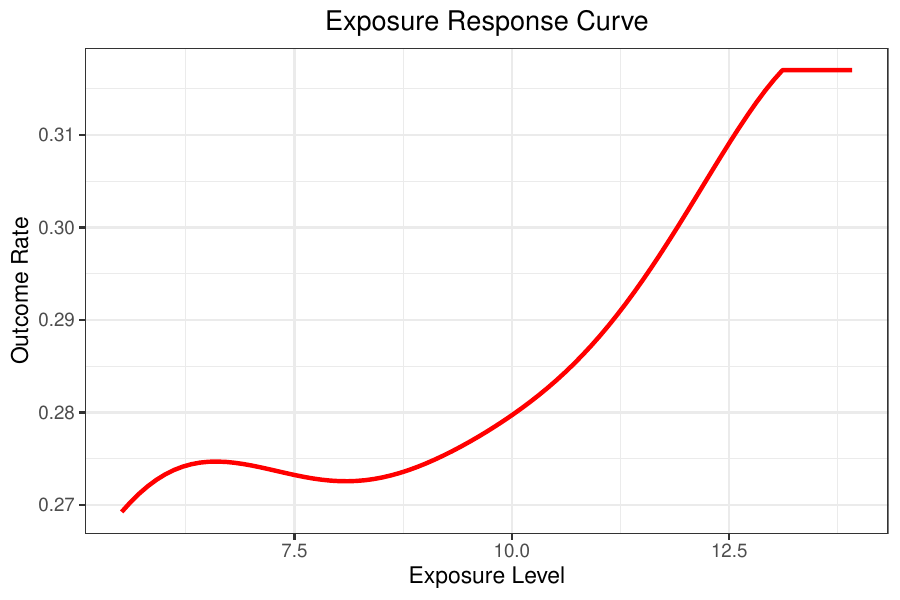}
    \caption{The causal ERF relating the percentage of people over 65 years old who did not finish high school to long-term \PM\ exposure estimated by a local polynomial regression model. \textbf{Note that the results are just for illustration purposes.}}
    \label{fig:example_1_erf}
\end{figure}

After achieving the acceptable covariate balance for \code{matching_3}, the matched data set will be used to estimate the ERF.  We use a non-parametric approach (\fct{estimate\textunderscore npmetric\textunderscore erf}) to fit the outcome model. The confidence intervals of the ERF can be obtained by a m-out-of-n bootstrap procedure proposed by \cite{wu2018matching}, which requires repeated implementation of the function \fct{estimate\textunderscore npmetric\textunderscore erf} on bootstrapped data sets. We omit the details of the confidence interval calculation, given the procedure is relatively straightforward yet time-consuming, while referring interested readers to see \cite{wu2018matching} for a real data example. Figure~\ref{fig:example_1_erf} shows the ERF obtained from the matched data set for \code{matching_3}.
\begin{CodeChunk}
\begin{CodeInput}
R> quant <- quantile(data$pm25, probs = c(0.1, 0.90))
R> set.seed(290)
R> erf_obj <- estimate_npmetric_erf(data$education,
                                    data$pm25,
                                    data$counter_weight,
                                    bw_seq = bw_seq=c(0.2, 1, 0.1),
                                    w_vals = seq(quant[1], quant[2], 0.1),
                                    nthread = 10,
                                    kernel_appr = "kernsmooth")
R> plot(erf_obj)
\end{CodeInput}
\end{CodeChunk}  

%% file: conclusion.tex
\section{Conclusion and future development} \label{sec:conclusion}

We have introduced the \pkg{CausalGPS} \proglang{R} package as a generic solution for causal inference with continuous exposures. Causal inference with continuous exposure has extended the research horizons of many scientific applications in social, economic, health, and medical research. Many studies in these scientific domains focus on understanding the causal effects of the duration of the exposure, the intensity of the exposure, or the dosage of the treatment, which is regarded as a continuous variable in nature. The \pkg{CausalGPS} package provides an essential causal inference tool for answering these scientific questions.
We also briefly summarized the causal inference workflow, which serves as the central backbone for our software development.
The main advantage of the causal inference workflow implemented by the \pkg{CausalGPS} package is the separation of the design and analysis phases, improving the objectiveness of causal inference. We have illustrated the matching approach using synthetic data, while the weighting approach can be implemented analogously. 

The package is computationally optimized for shared memory systems using \proglang{R}'s \pkg{parallel} package or \pkg{Rcpp}'s OpenMP support. The package can be extended in several future directions from implementation perspectives. In future software development, we plan to implement features to provide additional user specifications of the matching algorithm. This includes expanding the currently available nearest-neighbor matching (selecting the best-matched sample) to the k-nearest-neighbor matching (selecting the k best-matched samples), increasing the robustness of the matching algorithm to outliers. In addition, we may add new features, such as giving priority to match data samples from preferred groups, called preferential matching, which is helpful for analyzing clustered observational data \citep{arpino2016propensity}. 
From a computational standpoint, to accommodate big data, we intend to transition intensive computations to distributed memory systems and incorporate features that enable the exploration of parameter domains using optimization algorithms, such as genetic algorithms.

The package is available on CRAN and GitHub. Contributions from the open-source community, whether through opening issues, participating in discussions, or adding new features, are warmly encouraged.

%% file: computational_details.tex
\section{Computational details} \label{computational_details}

The computations were executed on a macOS system with the following specifications: a 2.6 GHz 6-core Intel Core i7 processor and 16GB of RAM, running on macOS Ventura 13.5.2. Docker was employed with an image built upon the rocker/verse:4.2.3 base. The Docker daemon was configured to utilize a substantial portion of the system's resources: 10 CPUs, 12GB of RAM, and a swap space of 4GB. A Docker image of our software is readily accessible on Docker Hub under the name \texttt{nsaphsoftware/causalgps\_paper}. It can be fetched directly or viewed online at \href{https://hub.docker.com/r/nsaphsoftware/causalgps_paper}{https://hub.docker.com/r/nsaphsoftware/causalgps\_paper}. For those interested in reproducing the results, the necessary code is available at \href{https://github.com/NSAPH-Software/CausalGPS/tree/master/functional_tests}{https://github.com/NSAPH-Software/CausalGPS/tree/master/functional\textunderscore tests} in the \texttt{functional\_tests} folder of the CausalGPS GitHub repository.


%% file: acknowledgments.tex
\section*{Acknowledgments}

Funding was provided by the HEI grant 4953-RFA14-3/16-4, US EPA grant 83587201-0, NIH grants R01 ES026217, R01 MD012769, R01 ES028033, R01 ES028033-S1, 1R01 ES030616, 1R01 AG066793-01R01, 3R01 AG066793-02S1, 1R01 ES029950, 1RF1 AG071024, 1RF1 AG074372-01A1, 2P30 ES009089-25, Alfred P. Sloan Foundation grant G-2020-13946, Harvard University Climate Change Solutions Fund, and Fernholz Foundation.
The authors report there are no competing interests to declare.


%% file: appendix.tex
\begin{appendix}



\end{appendix}